# Structure of A Heterogeneous Two-Phase Rotating Detonation Wave with Ethanol-Hydrogen-Air Mixture

Songbai Yao[a,b*], Xinmeng Tang[c], Wenwu Zhang[a,b]

[a]Zhejiang Key Laboratory of Aero Engine Extreme Manufacturing Technology, Ningbo Institute of Materials Technology and Engineering, Chinese Academy of Sciences, Ningbo 315201, China
[b]University of Chinese Academy of Sciences, Beijing 100049, China
[c]International Innovation Center of Tsinghua University, Shanghai 200062, China

**ABSTRACT**

In this short Letter, the structure of a rotating detonation wave (RDW) fueled by biofuel is revealed and expounded. The simulation is carried out under an Eulerian-Lagrangian framework in which the main characteristics of the two-phase RDW are analyzed in detail. Results suggest a self-sustained rotating detonation fueled by liquid ethanol and air can be achieved with hydrogen addition for combustion enhancement, and a laminated dual-front structure of the RDW due to the effect of droplet evaporation is captured and clarified.

*Corresponding author: yaosongbai@nimte.ac.cn (S. Yao)

# LETTERS

Rotating detonation engines (RDEs) fueled by liquid propellants are attracting widespread interest in view of the advance towards engineering applications, i.e., liquid fuels demonstrate advantages in terms of storage, transport, and combustion control. However, the research on liquid-fueled RDEs is still at the primary stage [1, 2]. Some early attempts [3, 4] were carried out for the RDE fueled by kerosene, and experimental and numerical studies [5-10] have been conducted to investigate the physics of two-phase rotating detonation waves (RDWs).

As a leading biofuel, ethanol is benefited by various characteristics such as low NOx and carbon monoxide, compared with conventional fossil fuels. In the latest studies, Yoneyama et al. [11] and Sato et al. [12] have carried out the first attempts to verify the feasibility of rotating detonations fueled by liquid ethanol and oxygen, in which it was also reported that the RDE could achieve a high specific impulse at various mass flow rates and equivalence ratios. Until recently, however, there has been little research on the study of the ethanol-fueled RDE. Therefore, in the present work, a numerical simulation of the RDE (see Fig. 1) fueled by liquid ethanol is conducted based on an Eulerian-Lagrangian framework with the aim to provide an insight into the structure of the two-phase rotating detonation wave (RDW).

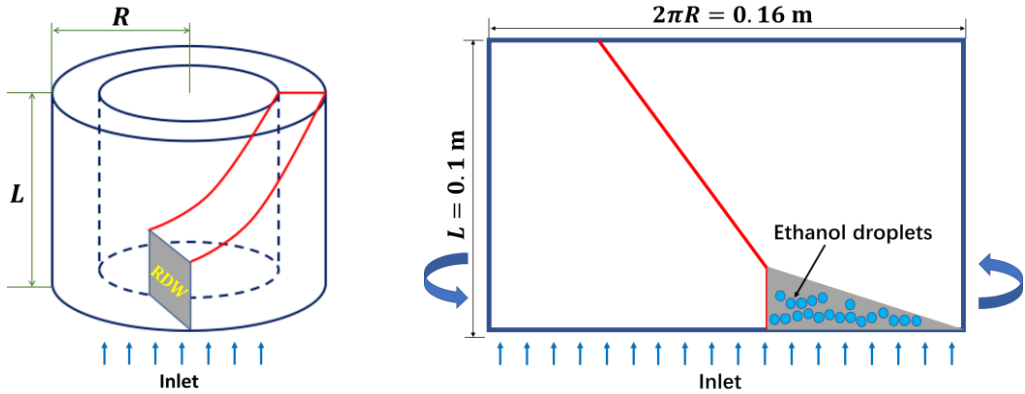

Figure 1. Schematic of the RDE and computation domain of an unwrapped combustor.

In this study, we solve the two-dimensional compressible Navier-Stokes equations of the carrier phase (gas and vapor), and the Lagrangian droplets are coupled through the source terms of mass, momentum, and energy equations. The chemical kinetics of ethanol and hydrogen oxidations are modeled by the one-step global mechanisms where the reaction rates $k_i$ in Arrhenius formulations are given by Westbrook and Dryer [13] and Marinov [14], respectively.

$$C2H5OH + 3O_2 \overset{k1}{\Rightarrow} 2CO2 + 3H_2O \qquad (1)$$

$$H_2 + 0.5O_2 \overset{k2}{\Rightarrow} H_2O \qquad (2)$$

The simulation is conducted using our developed solver based on the open-source rhoCentralFoam [15], which uses the central-upwind Kurganov and Tadmor schemes for shock capturing. The validity of rhoCentralFoam for detonation simulations has been widely tested, e.g., refs. [6, 16-19]. The Lagrangian particle tracking library is implemented and the transport equations for reacting species are solved. The thermal conductivity is computed according to the Eucken correlation [20] and the mass diffusivity is obtained under the relation of unity Lewis number. The carrier phase (droplets) is coupled with the gaseous phase through the source terms for the exchange of mass, momentum, energy and transport of species under the point-source assumption.

The Stokesian drag force is computed according to ref. [21] and the convective heat transfer is given by the model of Ranz and Marshall [22, 23]. In the present simulations, the droplets are assumed to be dilute (see Fig. 2) and the thus break-up process and the interactions between droplets are ignored. Readers are referred to our previous study [24] for the full description of the governing equations and numerical methods.

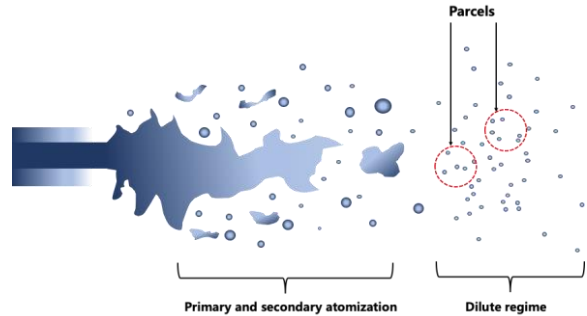

Figure 2. Numerical treatments of the injected droplets. Droplets are assumed to be within the dilute regime. A group of liquid droplets with constant mass and diameter is represented as a droplet parcel. Diagram redrawn based on Jenny et al. [25]

The inlet conditions are assumed to be uniformly-distributed micro nozzle flow, which is very common treatment to approximate the fuel inlet for RDE simulations [6, 17, 18, 26]. Liquid droplets will be injected from the inlet at the same velocity of the gaseous phase. Adiabatic, slip and non-catalytic boundary conditions are enforced on the outer wall. A wave transmissive non-reflecting boundary is applied to the outlet where the far-field condition is set to the ambient temperature and pressure of $p_{\text{inf}} = 1.013 \times 10^5$ Pa and $T_{\text{inf}} = 300$ K, respectively.

The computational domain (160 mm $\times$ 100 mm) in Fig. 1 is discretized in a two-dimensional Cartesian coordinate system, i.e., the rotating detonation chamber (RDC) is unwrapped along the circumferential direction along which the mesh cells are spaced uniformly at $\Delta x = 100$ μm. On the other hand, the mesh cells are stretched downstream along the y direction by a factor of 10, which will lead to a minimum size of $\Delta y = 50$ μm near the head end of the RDC to ensure high resolution of the wave front and a larger mesh size at the outlet where the flow field is mainly composed of burned products and low-resolution mesh will suffice. Also, an error accumulation analysis using the proposed method [27] indicates a total error $S_{\text{err}} < 0.1\%$, well within allowable ranges.

The RDE is ignited by a premixed gaseous hydrogen-air mixture at the stoichiometric concentrations at an inlet total pressure and temperature of 2 MPa and 300 K, respectively, an approach that was used in the experiment to ensure ignition of liquid-fueled detonations [28]. Starting from the hydrogen-air RDE in a stable state (denoted as $t_0$ for convenience), liquid droplets at a diameter of 10 μm will start to be injected into the flow from the inlet, and the air stream is also supplied. Instead of using pure oxygen for combustion enhancement [11], here liquid ethanol will mix with air oxidizer, and thus a small amount of hydrogen is added to improve the ignitability and reactivity of the reactant mixture [3, 4, 29]. The volume fraction of hydrogen is 17.4% in the gaseous stream. Noted that it is possible to reduce the amount of hydrogen addition, but the lower limit requires further analysis. Unless otherwise stated, the gaseous air stream with hydrogen addition will be hereinafter referred to as the air stream. The initial temperature of the liquid ethanol droplets is set to the ambient temperature, i.e., the droplets are not pre-heated. The inlet total pressure $p_0$ is fixed at 2.0 MPa. To enhance the evaporation of liquid droplets, the hot air stream is injected at a total temperature of $T_0$ = 600K. The mass flow rates of liquid ethanol and air stream are 0.05 kg/s and 0.50 kg/s, respectively, corresponding to a global equivalence ratio of $\phi \approx 1.3$.

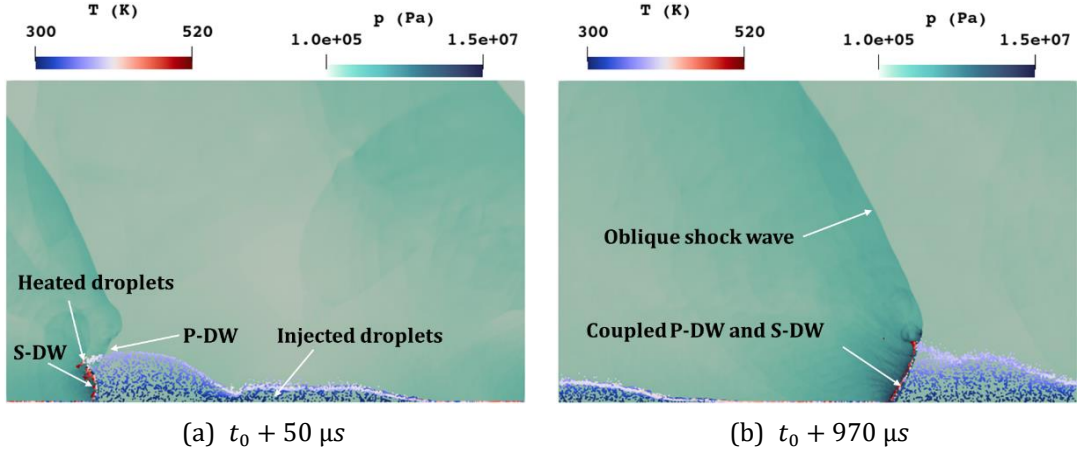

(a) $t_0 + 50$ μs  (b) $t_0 + 970$ μs

Figure 3. Pressure distributions of the flow field. Initial and stable stages of the two-phase RDW after the supply of liquid ethanol droplets and air stream (with 17.4% $H_2$). Liquid droplets are colored by temperature.

Shortly after the injection of droplets for 50 μs, a cluster of liquid ethanol droplet can be found in front the of the detonation wave (see Fig. 3a). As the liquid ethanol is injected at an ambient temperature, after the droplets mix with the hot air, the evaporation starts to progress gradually. When the droplets pass through the preceding detonation wave (P-DW), they are instantly heated by the wave front that results in significant temperature rise and fast evaporation. On the other hand, the local temperature of the gaseous products behind the detonation wave is cooled due to the latent heat of evaporation of the droplets. The liquid droplets will be heated to a further extent after mixing with the hot products in the expansion region and ethanol vapor starts to accumulate, which results in the onset of a succeeding detonation wave (S-DW) after the P-DW. Though the formation of the secondary DW was similar to that reported by Ren et al. [7], there is key difference: the latter was wave bifurcation caused by a curved secondary shock originating from the injection plate, whereas the S-DW in the current study will develop and eventually lead to a laminated structure, as will be discussed later. The emerging S-DW is supported by the consumption of ethanol vapor and starts to gain strength during propagation. After several propagation cycles, a stable self-sustained RDW structure is achieved (see Fig. 3b). In the current structure, the P-DW serves as the driving energy to pre-heat the injected liquid droplets at lower temperatures, while the S-DW is maintained at the supply of ethanol vapor and air mixture, and pushes forward the P-DW in return. The P-DW and S-DW are fully coupled during stable propagation at an estimated speed of 1501 m/s, indicating a velocity deficit of about 22.5% compared with the theoretical Chapman–Jouguet (C-J) value [30]. The extent of the detonation velocity deficit is in accordance with the experimental research on two-phase RDWs in which a deficit of 20-25% was reported [3, 31].

Here we present a detailed analysis of the composition of the two-phase RDW. In Fig. 4a, the ethanol vapor and evaporating droplets are shown where the contact surface is also annotated. Similar to the structure of gaseous RDWs, the contact surface will separate the burned product from current and previous cycles; shearing vortices will form along the contact surface due to the velocity deficits of the burned products that result in Kelvin-Helmholtz instability on the interface. However, as can be seen from the distribution of the ethanol vapor, there is noticeable amount of unreacted ethanol on the contact surface, meaning that not only burned products but also unburned gaseous ethanol vapor exists.

To expound the laminated structure of the two-phase RDW, a close-up view of the heat release rates of the reacting fronts is also provided in Fig. 4a. In stable propagation, the P-DW and S-DW maintain a distance and create an evaporation zone where droplets pre-heated by the P-DW will reside until they are fully vaporized and detonated by the S-DW. The distance between the P-DW and S-DW maximizes near the inlet and gets shorter downstream; eventually, the P-DW and S-DW connect at the contacting point C. This is because downstream liquid droplets in the triangular-shaped fuel refill zone have already been heated by the hot air stream to higher temperatures; therefore, they will be vaporized and detonated instantly after being swept by the DWs. In summary, a schematic diagram is shown in Fig. 4b to demonstrate the laminated structure of the two-phase RDW. Noted that this laminated structure is closely related to the effect of droplet evaporation in of heterogenous mixture consisting of gaseous hydrogen and liquid ethanol. For a gaseous ethanol (vapor)-air mixture, the structure of the RDW will not exhibit such features.

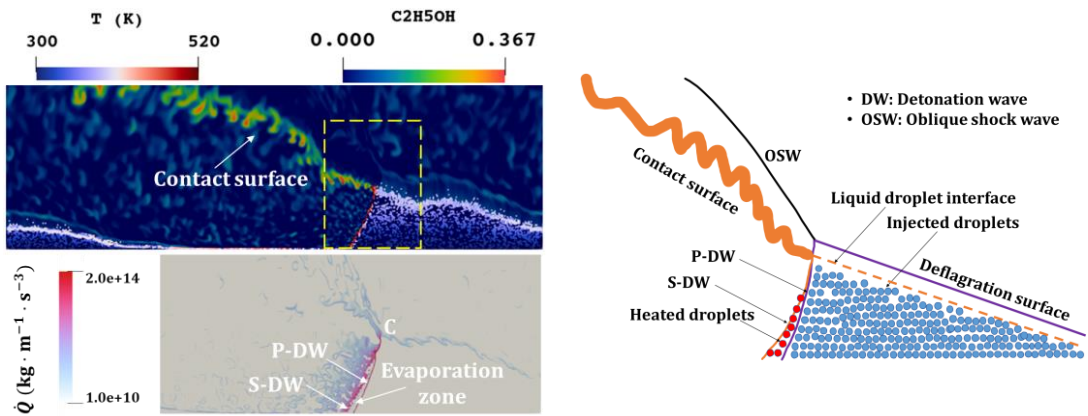

(a) Ethanol vapor and evaporating droplets. Heat release rates of the coupled DWs.

(b) Schematic of the laminated structure

Figure 4. Laminated structure of the two-phase RDW during stable propagation.

Additionally, a grid sensitivity analysis is provided here for validations. The simulations are compared at three different resolutions where the total numbers of grid nodes for Mesh-1, Mesh-2, and Mesh-3 are $800 \times 250$, $1600 \times 500$, and $3200 \times 500$, respectively. Results are shown in Fig. 5 in which the snapshots of the temperature and Mach number contours are presented. It is found that all simulation cases converge and there is no failure or chaotic propagation of RDWs caused by the change of resolution. Under a finer mesh ($3200 \times 500$), more details of the contact surface can be captured and the wavy-shaped deflagration surface is clearer; but the main structure of the two-phase RDW and the distributions of the evaporating droplets are approximately the same under the three different resolutions. In consideration of both computational efficiency and accuracy, Mesh-2 has been implemented in the present study.

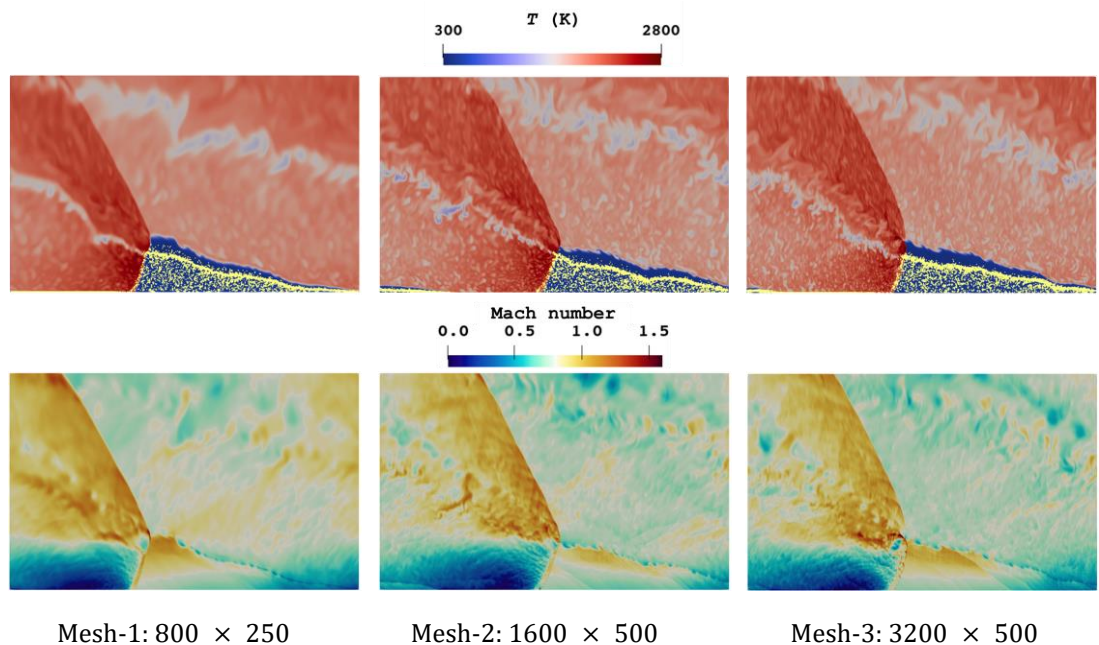

Mesh-1: 800 × 250     Mesh-2: 1600 × 500     Mesh-3: 3200 × 500

Figure 5. Simulation cases of two-phase RDWs under various resolutions.

In conclusion, we reveal the main characteristics of the structure of two-phase RDW fueled by a heterogenous ethanol-hydrogen-air mixture. A numerical study based on the Eulerian-Lagrangian framework is conducted where liquid droplets are injected at an ambient temperature and mix with the hot air stream with an addition of hydrogen for combustion enhancement. For the self-sustained RDW, it is found that there exists a laminated structure where droplets will be sandwiched in between: the preceding detonation wave will provide the latent heat of vaporization required for the first-stage evaporation of the droplets, and then the pre-heated droplets will vaporize and be detonated by the succeeding detonation wave. This laminated structure will remain fully coupled during the stable propagation of the RDW.

**ACKNOWLEDGMENTS**

S. Yao acknowledges the support from the Chinese Academy of Sciences and Ningbo Yongjiang Talent Introduction Programme (No. 2022A-210-G).